# Bandgap Controlling of the Oxygen-Vacancy-Induced Two-Dimensional Electron Gas in $SrTiO_3$

Z. Q. Liu[1,2#]*, W. Lu[3], S. W. Zeng[1,2], J. W. Deng[1], Z. Huang[1], C. J. Li[1,4], M. Motapothula[1,5], W. M. Lü[1], L. Sun[1], K. Han[1,2], J. Q. Zhong[2], P. Yang[6], N. N. Bao[1],W. Chen[2], J. S. Chen[3], Y.P. Feng[2], J. M. D. Coey[1,7], T. Venkatesan[1,2,4,8], and Ariando[1,2]*

[1]*NUSNNI-Nanocore, National University of Singapore, 117411 Singapore*

[2]*Department of Physics, National University of Singapore, 117542 Singapore*

[3]*Department of Materials Science and Engineering, National University of Singapore, 117576 Singapore*

[4]*National University of Singapore (NUS) Graduate School for Integrative Sciences and Engineering, 28 Medical Drive, Singapore 117456*

[5]*Center for Ion Beam Applications, National University of Singapore, 117542 Singapore*

[6]*Singapore Synchrotron Light Source (SSLS), National University of Singapore, 5 Research Link, Singapore 117603*

[7]*Department of Pure and Applied Physics, Trinity College, Dublin 2, Ireland*

[8]*Department of Electrical and Computer Engineering, National University of Singapore, 117576 Singapore*

[#]Current address: Oak Ridge National Laboratory, Oak Ridge, Tennessee 37831, United States.

*To whom correspondence should be addressed.
E-mail: liuz@ornl.gov
E-mail: ariando@nus.edu.sg

**Strongly correlated oxides are full of fascinating phenomena owing to their interacting lattice, charge, spin and orbital degrees of freedom.[1–5] Bandgap, a critical parameter for an oxide insulator, is well determined by those degrees of freedom and in turn directly affects electronic, magnetic and optical properties of the material. Typically, tunability of the bandgap in an oxide insulator can be achieved through chemical doping,[6] which is important for electronic and photonic device applications. Here we report large bandgap enhancement in $SrTiO_3$ (STO) thin films, which can be up to 20% greater than the bulk value, depending on the deposition temperature. There is no significant change in density and cationic ratio of the oxide so the effect is attributed to Sr/Ti antisite defects, an attribution supported by density functional theory calculations. It was found that the bandgap enhancement significantly changes the electronic and magnetic phases in the oxygen-vacancy-induced two-dimensional electron gas at the interface between amorphous $LaAlO_3$ (LAO) and STO. This opens an attractive path to tailor electronic, magnetic and optical properties of STO-based oxide interface systems under intensive focus in the oxide electronics community. Meanwhile, our study provides key insight into the origin of the fundamental issue that STO thin films are difficult to convert into metals by oxygen vacancy doping.**

STO is central to modern oxide electronics since it serves as the main workhorse for complex functional oxide heterostructure fabrications. After the two-dimensional electron gas (2DEG) at the interface between STO and LAO had been unveiled,[7] a large number of exotic properties of the 2DEG were revealed such as a critical thickness for the appearance of conductivity,[8] Kondo effect,[9] interface superconductivity,[10] an electrically tunable ground state,[11,12] electronic phase separation[5] and recently discovered high-temperature superconductor-like gap behavior.[13] In addition, the practicability of monolithically integrated oxide electronics based on the LAO/STO interface system has been demonstrated.[14]

As the research on the LAO/STO interface system is going on, the properties of STO itself have become a center of attention. For example, oxygen vacancies in STO have been a longstanding issue of debate in understanding the emergence of novel electronic and magnetic phases in the 2DEG. Additionally, a fundamental issue related to STO remaining up to now is that although STO single crystals are easy to make metallic with oxygen vacancies,[15,16] it is rather difficult to make an insulating STO thin film completely metallic via oxygen vacancies. Instead, oxygen-deficient STO films are typically semiconducting at low temperatures.[17–20] Besides, the 2DEG at the LAO/STO interface fabricated on STO films[21–24] is more localized at low temperatures than that fabricated on STO single crystals;[7] it was found that the insertion of a STO film layer degrades the LAO/STO interface significantly;[25,26] the fully metallic state of the LAO/STO interface based on STO films is achieved when STO films are deposited at a very high temperature of 1100 °C.[27] These perhaps indicate the presence of point defects/disorder in the STO films deposited at the typical temperature range of 600-800 °C.

A 2DEG in STO-based heterostructures can be created via oxygen vacancies (2DEG-V) when the overlayer contains elements such as Al which have a strong affinity for oxygen.[28–30] Hence the 2DEG-V can be easy to realize by depositing an amorphous, nonpolar LAO (aLAO) film onto STO at room temperature. In this work, our focus is on the 2DEG-V at the interface between amorphous LAO films and STO films, and its relation to the bandgap of STO films, which was found to significantly increase due to the defective nature of STO thin films.

In this work, we fabricated 2DEG-V heterostructures by depositing aLAO films on STO films that were pre-deposited on LAO single-crystal substrates. STO films of 150 nm thickness were deposited from a single-crystal STO target on (100)-oriented LAO single-crystal substrates, using pulsed laser deposition (KrF laser $\lambda$ = 248 nm) in $10^{-2}$ Torr oxygen partial pressure at various temperatures ranging from 30 to 750 °C. After deposition, the STO films were cooled to room temperature in the deposition oxygen pressure. Amorphous 25-nm-thick LAO films were subsequently deposited from a single-crystal LAO target on top of STO films in $10^{-6}$ Torr oxygen pressure at room temperature. During all depositions, the laser fluence was fixed at 1.8 $J/cm^2$ and the repetition rate was 5 Hz.

The structure of aLAO/STO/LAO heterostructures is schematized in **Figure 1**(a). Depending on deposition temperature, the STO layer can be either amorphous or crystalline. Figure 1(b) shows a cross-section transmission electron microscopy (TEM) image of an aLAO/STO/LAO heterostructure where the 150-nm-thick STO layer was deposited at 750 °C. It covers the LAO substrate uniformly and the interface with the aLAO cap layer is reasonably flat. The zoom-in image in Figure 1(c) confirms the amorphous nature of the LAO capping layer and the crystalline nature of the underlying STO film.

The STO films grown in $10^{-2}$ Torr at temperatures ranging from 30 to 400 °C are predominantly amorphous as no x-ray diffraction peak of STO was detected (see Figure S1 of the Supporting Information). At 450 °C weak STO diffraction peaks start to appear, which increase in intensity as the deposition temperature is raised. All STO films deposited at relatively high oxygen partial pressure ($>10^{-5}$ Torr) were found to be highly insulating (resistance > GΩ). No $Ti^{3+}$ was detected in x-ray photoelectron spectroscopy (XPS) measurements (see Figure S2 of the Supporting Information). The interface between an STO film and a LAO single crystal substrate is known to be insulating[31] as well. Measurable room-temperature conduction only appeared upon depositing aLAO on top of STO films.

The STO-growth-temperature dependence of room-temperature sheet resistance of aLAO/STO/LAO heterostructures is shown in **Figure 2**(a). Below 400 °C, the STO films are amorphous and the heterostructures are insulating. As the STO films start to become crystalline from 450 °C, the heterostructures begin to exhibit measurable conductivity, with room-temperature sheet resistance of the order of $10^4$ to $10^5$ Ohm per square. This indicates 2DEG-V formation at the interface between aLAO and STO. Temperature-dependent sheet resistance measurements were performed for the heterostructures with crystalline STO [Figure 2(b)].The aLAO/STO/LAO heterostructure with the STO film deposited at 450 °C is insulating, while those with STO films fabricated at 600 and 750 °C present a large resistance upturn at 148 and 132 K, respectively. In contrast, the aLAO/STO heterostructure produced by depositing a 25-nm-thick aLAO layer on top of an STO single-crystal substrate at room temperature and $10^{-6}$ Torr oxygen pressure is much more conductive and metallic over the entire temperature range.

The temperature-dependent sheet carrier density and mobility of aLAO/STO/LAO heterostructures with crystalline STO layers and the aLAO/STO heterostructure are illustrated in Figure 2(c). All the conductive heterostructures show the carrier freeze-out effect—a pronounced decrease in carrier density with lowering temperature, which is the signature of the 2DEG-V.[30] The sheet carrier density of aLAO/STO/LAO heterostructures with STO films fabricated at 600 and 750 °C is $9.36 \times 10^{13}$ and $7.54 \times 10^{13}$ $cm^{-2}$ at room temperature, falling to $3.61\times 10^{13}$ and $2.89 \times 10^{13}$ $cm^{-2}$ at 10 K, respectively. The temperature-dependent mobility of the aLAO/STO heterostructure [Figure 2(d)] is similar to that of reduced STO single crystals,[15] where the mobility falls exponentially with increasing temperature above 30 K due to phonon scattering and further increases to ~700 $cm^2/(V\cdot s)$ at 10 K as a consequence of the dielectric screening of ionized scattering potentials. On the contrary, the 2DEG-V built on STO films has a much smaller mobility between 0.8 and 7.5 $cm^2/(V\cdot s)$, and it shows a mobility downturn at low temperatures. These features clearly demonstrate the localization of electrons at low temperatures in the STO-film-based 2DEG-V.

To figure out the difference between the 2DEG-V built on STO films and that fabricated on STO single crystals, we measured ultraviolet-visible-infrared (UV) transmittance for 150-nm-thick STO films deposited on LAO single-crystal substrates at different temperatures. The large bandgap of the substrate (5.6 eV) enables us to measure transmittance spectra of STO films down to 220 nm (**Figure 3**). The UV spectrum of a STO single crystal substrate is shown as well for reference. Compared with STO single crystals, the UV spectra of all our STO films exhibit a large blue shift. The fitted bandgap of amorphous STO films deposited between 30 and 400 °C reaches 3.95 eV, 0.70 eV larger than that an STO crystal. The crystalline STO films fabricated at 450, 600 and 750 °C have a fitted indirect optical bandgap of 3.75, 3.59 and 3.56 eV, respectively. The quantum size effect in 150-nm-thick STO films is negligible as the energy scale due to the dimension confinement is ~1 meV, which is two orders of magnitude smaller than the bandgap enhancement in crystalline STO films.

Cationic vacancies could lead to the bandgap enhancement because the O 2*p* band is no longer full and meanwhile the crystal field splitting of Ti atoms in the oxygen octahedron is reduced in the present of cationic vacancies (see Figure S3 of the Supporting Information). In this case, an estimated 6% cationic vacancies is expected to generate a bandgap increase of 0.35 eV in crystalline STO films. This corresponds to a reduction in density of 4.4%. We therefore performed x-ray reflectivity for our crystalline STO films. But, we found that the density of a c-STO film obtained by synchrotron-based x-ray reflectivity[32] was 5.16±0.08 g/cm$^3$ (see Figure S4 of the Supporting Information), comparable with 5.11 g/cm$^3$ for an ideal STO single crystal. In addition, cationic vacancies in oxide films are expected to be more in high oxygen pressure depositions[33] and STO films deposited at various oxygen partial pressures ($10^{-2}$~$10^{-5}$ Torr) exhibit similar bandgap (see Figure S5 of the Supporting Information), which further supports that the bandgap enhancement is not due to cationic vacancies.

The composition of the PLD-grown crystalline STO (c-STO) films was found to be stoichiometric in Rutherford backscattering spectrometry (RBS) experiments (see Figure S6 of the Supporting Information). This is quite consistent with the fact that the density of a c-STO film is comparable with the STO single crystal. It is known that the laser fluence could to some extent change the stoichiometry of oxide films in PLD.[34] We deposited STO films by different laser fluence ranging from 1.3-3.0 J/cm$^2$ and it was found that the bandgap of crystalline STO films is not sensitive to the laser fluence (see Figure S7 of the Supporting Information). This implies that the bandgap enhancement is not a result of the (non)stoichiometry issue. Furthermore, reciprocal lattice mapping reveals that the unit cell volume of the STO film is also comparable with that of an ideal STO single crystal (see Figure S8 of the Supporting Information).

For 150-nm-thick STO films deposited on LAO substrates at 750°C, the out-of-plane lattice constant was measured to be 3.909 Å. The value is quite close to the lattice constant 3.905 Å of bulk STO, which reveals that epitaxial strain has largely relaxed. Due to the large lattice mismatch between STO and LAO, misfit dislocations exist in STO films for strain relaxation. To examine the effect to lattice misfit on the STO bandgap, STO films were deposited on other large bandgap substrates (MgO and sapphire). However, it was found that such STO films have similar bandgap compared with those grown on LAO (see Figure S9 of the Supporting Information). This suggests that the strain and misfit dislocation is not the predominant origin of the bandgap enhancement.

So now how can we understand the large increase in bandgap in the STO films (Figure 3)? Here we have recourse to density functional theory (DFT) calculations. Due the lack of periodicity, a cluster model was used for the amorphous state. The calculated indirect bandgap for an ideal STO single crystal was ~1.83 eV while it is ~2.80 eV for the amorphous model (see Figure S10 and S11 of the Supporting Information). DFT calculations do not accurately reproduce the magnitude of the bandgap,[35] but they do indicate trends, so this is qualitatively consistent with our experimental observation.

Point defects are extensively present in c-STO films deposited at relative low temperature such as 600-800 °C compared with the STO single crystal growth temperature, which lead to significant dielectric losses.[36] The influence of cation vacancies and Sr/Ti antisite defects on the bandgap of c-STO was modeled in DFT calculations on a 2×2×2 cell. The removal of Sr and Ti atoms does not increase the bandgap (see Figure S12 of the Supporting Information), but a single Sr/Ti antisite defect produced a pronounced increase in the bandgap from 1.83 to 2.30 eV (**Figure 4**). Therefore, it appears that a certain concentration of antisite defects in the PLD-grown STO films is probably responsible for the bandgap enhancement. The bandgap would certainly affect the donor level of oxygen vacancies and typically as in conventional semiconductors the donor level becomes deeper when the bandgap is enlarged.[37] Also, the donor level of oxygen vacancies in a very large bandgap insulator such as LAO (~5.6 eV) is much deeper.[38,39] Therefore, the oxygen-vacancy-induced 2D conduction is more localized in STO-film-based 2DEG-V and that is also why the oxygen-vacancy-induced 3D conduction shows the low-temperature semiconducting phase in oxygen-deficient STO films.[17–20] Now looking back into the growth-temperature-dependent bandgap in Figure 3, we can draw a conclusion that a high growth temperature is useful to suppress point defect such as Sr/Ti defects. For example, the STO single crystal growth temperature is more than 1500 °C, which helps to minimize point defects.

To examine the effect of bandgap enhancement (or localization) on the magnetic state of the 2DEG-V, magnetotransport properties of the 2DEG-V were studied at low temperatures. The parallel magnetoresistance (MR), where the orbital effect is minimized, is predominantly negative in aLAO/STO/LAO heterostructures with STO films deposited at 750 °C (**Figure 5**). More importantly, as shown in the inset of Figure 5, the MR curves exhibit hysteresis in continuous field scans from 0 → 9 T → -9 T → 9 T. On the contrary, the parallel MR of the 2DEG-V fabricated on STO single crystals was predominantly positive and no hysteresis was observed (see Figure S13 of the Supporting Information). Hysteretic MR is an indication of magnetic order, which has been observed in crystalline LAO/STO heterostructures[9,40] and other STO-based systems.[41,42]

Anisotropic MR (AMR) of the STO-film-based 2DEG-V was also examined. **Figure 6**(a) demonstrates the AMR effect when a field of 9 T is rotated from out-of-plane (OP) to in-plane (IP). In this measurement geometry, magnetic field is always normal to current. Strong anisotropy between OP- and IP-MR demonstrates the two-dimensional signature of the conduction. Regardless of the field direction, the MR under 9 T is negative, suggesting the dominant role of the ferromagnetic order. IP-AMR was measured with a 9 T field as well [Figure 6(b)]. A four-fold oscillation in the IP-AMR can be clearly seen. Indeed, the four-fold oscillation in the IP-AMR has been recently observed in crystalline LAO/STO heterostructures.[43] It was found that the four-fold oscillation can only be seen in two-dimensional electron systems while the IP-AMR is two-fold for three-dimension conduction systems. Thus the emergence of the four-fold oscillation in our case again corroborates the 2D signature of conduction. Moreover, the origin of the four-fold oscillation was suggested to be magnetic scatterings from localized Ti $d_{xy}$ moments coupled to the cubic crystal symmetry in STO,[43] which is consistent with the ferromagnetic order revealed by the hysteretic MR.

Recent calculations by Pavlenko *et al*.[44] suggested that magnetism at the LAO/STO interface was not an intrinsic property of the 2DEG-P but resulted from the orbital reconstruction induced by oxygen vacancies.[45] On the other hand, the ferromagnetism at the LAO/STO interface has been consistently attributed to localized Ti 3*d* electrons,[5,9,46–50] which has been experimentally found to be due toTi $d_{xy}$ orbitals in the $t_{2g}$ band.[51] The oxygen-vacancy-induced orbital reconstruction at the LAO/STO interface has been experimentally observed by a recent resonant soft-x-ray scattering study,[52] where the degeneracy of Ti $t_{2g}$ orbitals was seen to be lifted by oxygen vacancies, with the energy of $d_{xy}$ orbital lowest. Therefore, the magnetic order observed in the 2DEG-V built on STO films in our case can be understood in the following way: the 2DEG-V formation induces orbital reconstruction of interface Ti $t_{2g}$ orbitals; as the 2DEG-V is largely localized, many electrons from oxygen vacancies are localized in $d_{xy}$ orbitals, which then order magnetically.

In conclusion, by integrating the STO-film-based interfacial 2DEG-V onto LAO substrates we have been able to tune its electronic properties in STO films with different bandgap. The electrons in the aLAO/STO/LAO heterostructures are always more localized than those at the interface of STO single crystals, and they are completely insulating when the STO is amorphous. A common cause of the localization and the bandgap enhancement in STO films has been identified as the bandgap enhancement in STO films due to Sr/Ti antisite disorder. Moreover, the existence of magnetic order in the STO-film-based 2DEG-V is inferred from magnetotransport measurements. Our work opens an attractive way to tailor the electronic, optical and magnetic properties in STO-based 2DEG systems under intensive focus in the oxide electronics community.

## Experimental Section

*Sample characterization*: The deposition rates of STO films and amorphous LAO (aLAO) films were calibrated by TEM measurements. Structural characterization was performed by TEM and x-ray diffractometry; chemical composition and valence were analyzed by RBS and XPS, respectively; electrical measurements were carried out in a Quantum Design physical property measurement system with electrical contacts onto $5 \times 5\ mm^2$ samples made with Al wires using wire bonding. While the Hall effect of all aLAO/STO/LAO heterostructures were measured in the Van der Pauw geometry, sheet resistance and MR measurements were conducted in the standard four-probe linear geometry. Transmittance spectra of STO films grown on LAO substrates were examined by UV spectroscopy.

## Acknowledgements

We thank the National Research Foundation (NRF) Singapore under the Competitive Research Program (CRP) “Tailoring Oxide Electronics by Atomic Control” (Grant No. NRF2008NRF-CRP002-024), the National University of Singapore (NUS) for a cross-faculty grant, and FRC (ARF Grant No.R-144-000-278-112) for financial support. PY is supported from SSLS via NUS Core Support C-380-003-003-001.

**Figure 1**

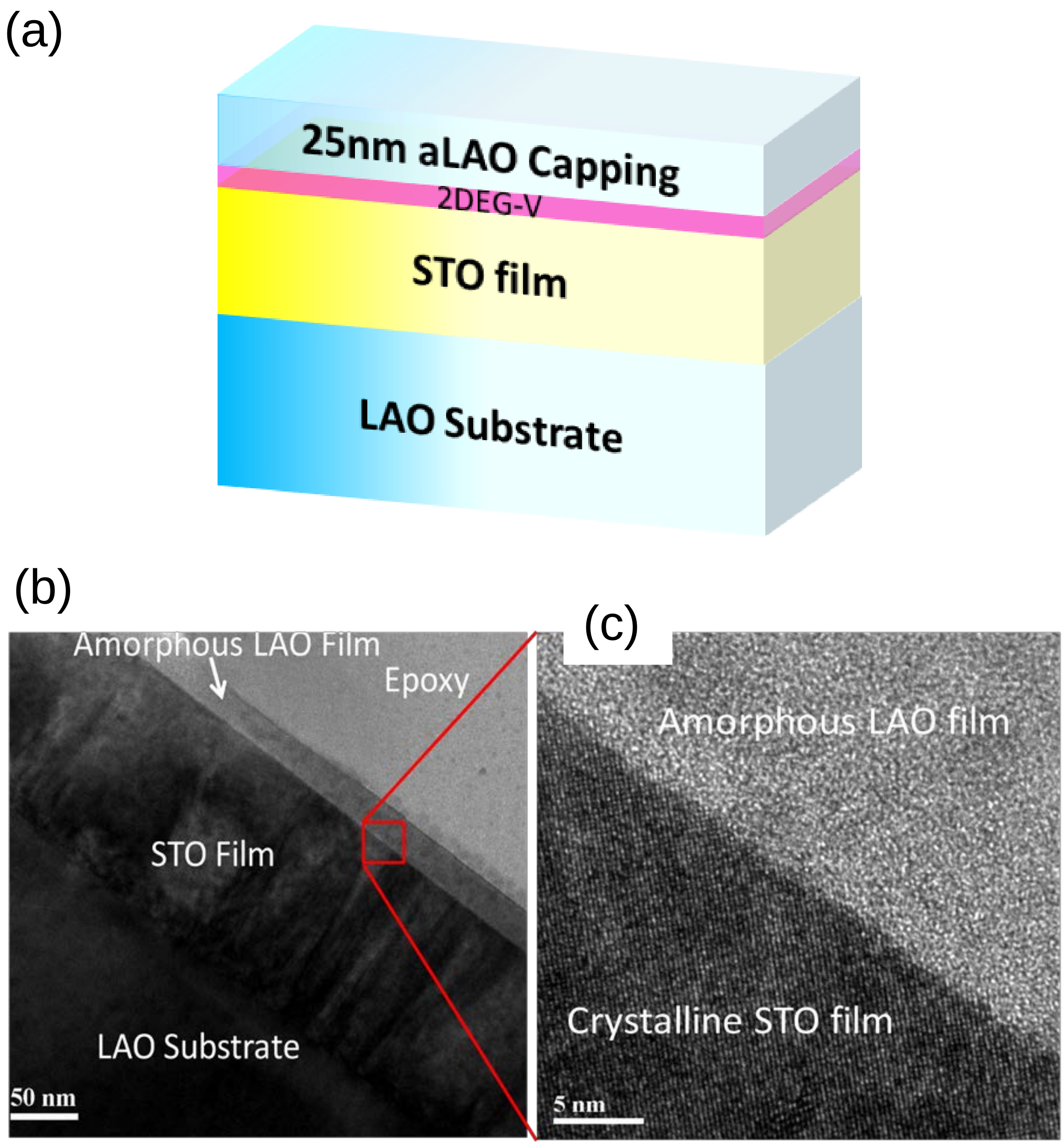

**Figure 1. Structural characterization.** (a) Schematic of aLAO/STO/LAO heterostructures. (b) Cross-section TEM image of an aLAO/STO/LAO heterostructure with the STO layer deposited at 750 °C. (c) Zoom-in image of an interface region marked by the red hollow square in (b).

**Figure 2**

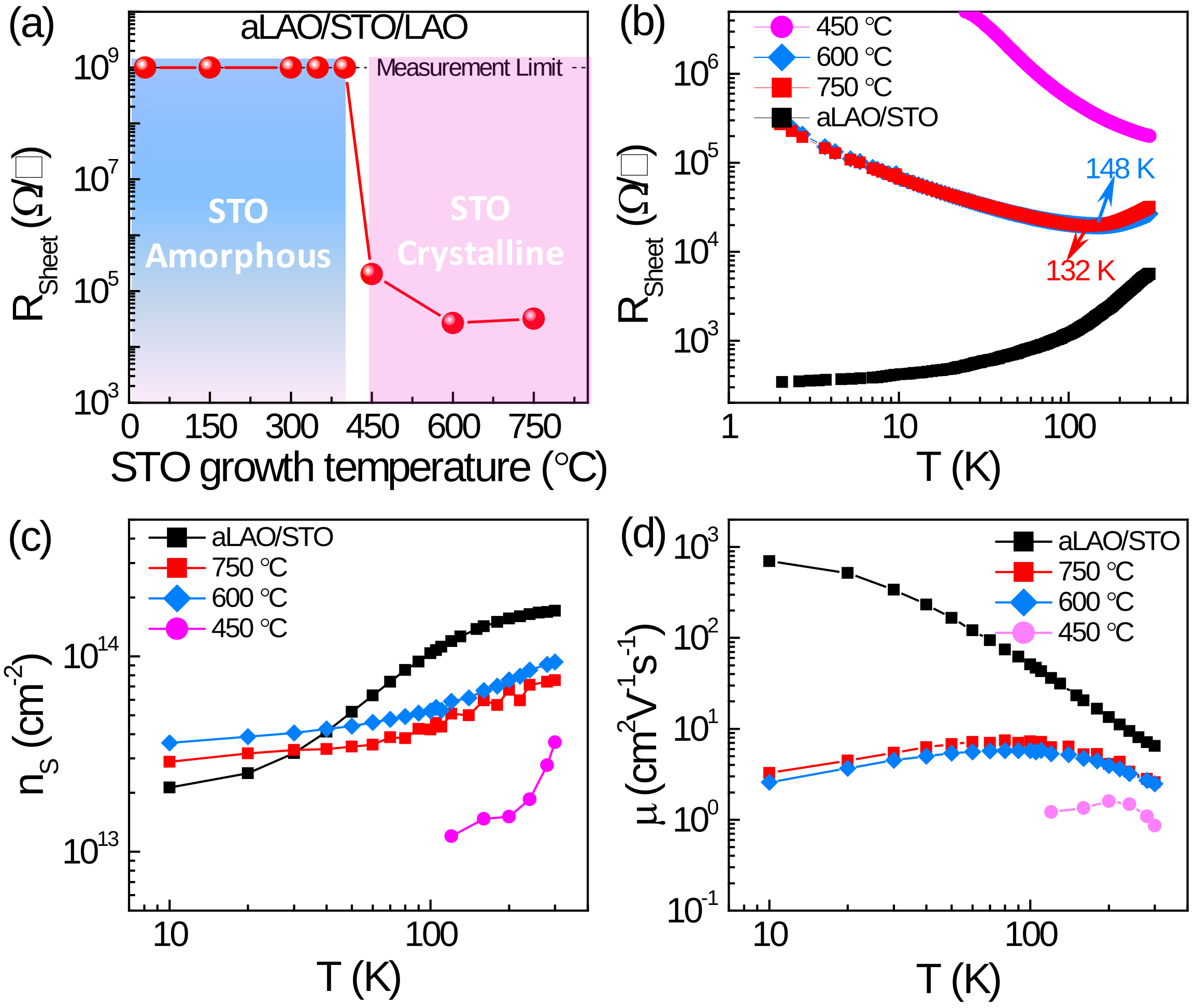

**Figure 2. Electrical transport properties.** (a) STO-growth-temperature dependence of room-temperature sheet resistance of aLAO/STO/LAO heterostructures. (b) Temperature-dependent sheet resistance of an aLAO/STO heterostructure with the aLAO layer deposited on a STO substrate at$10^{-6}$ Torr oxygen pressure at room temperature and aLAO/STO/LAO heterostructures with STO films deposited above 400˚C. The arrows indicate resistance upturn temperatures. Temperature dependence of sheet carrier density in (c) and mobility in (d) of the aLAO/STO heterostructure and aLAO/STO/LAO heterostructures with STO films deposited at different temperatures.

**Figure 3**

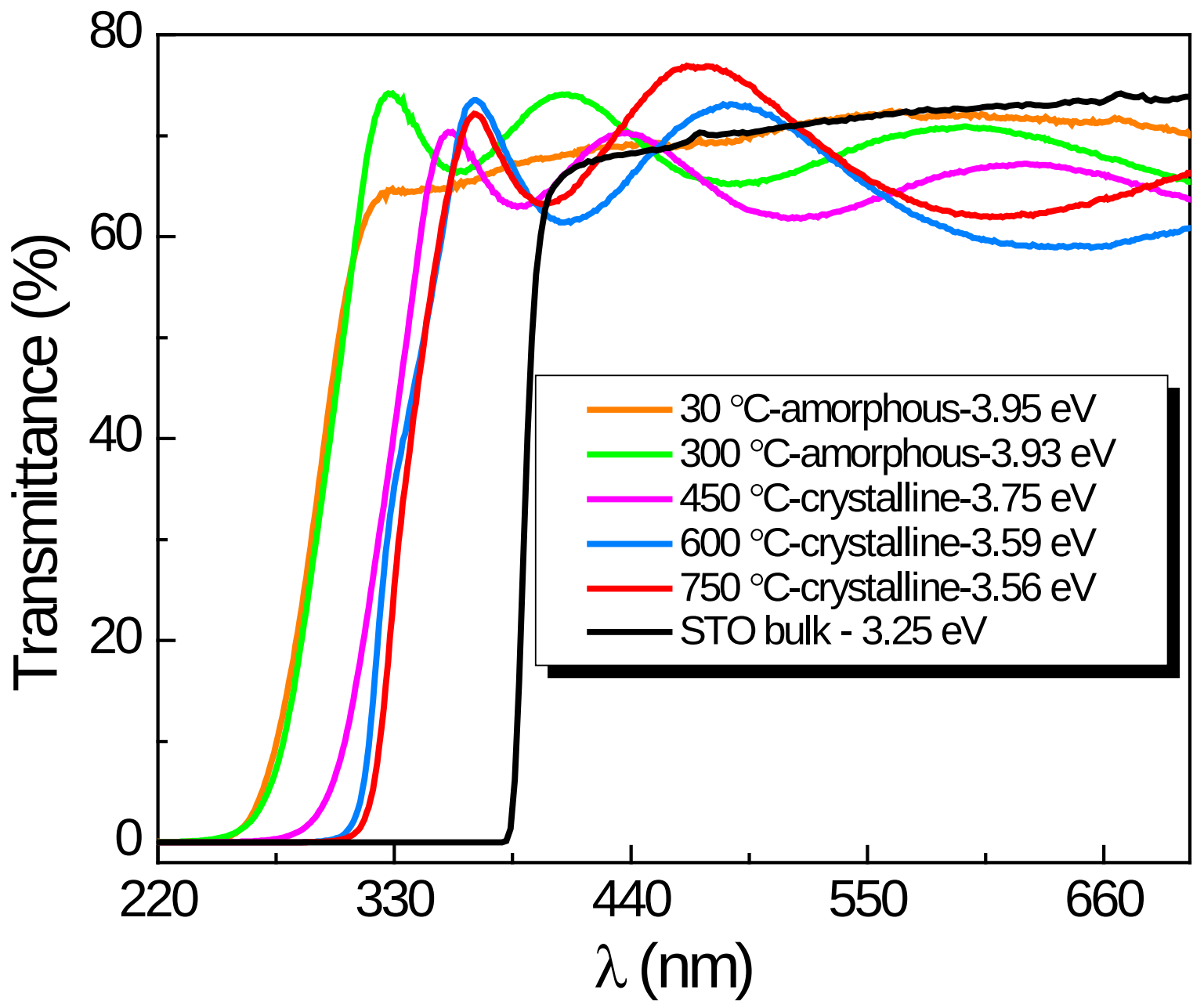

**Figure 3. UV spectra.** UV spectra of a STO single-crystal substrate and STO films deposited on LAO substrates at different temperatures.

**Figure 4**

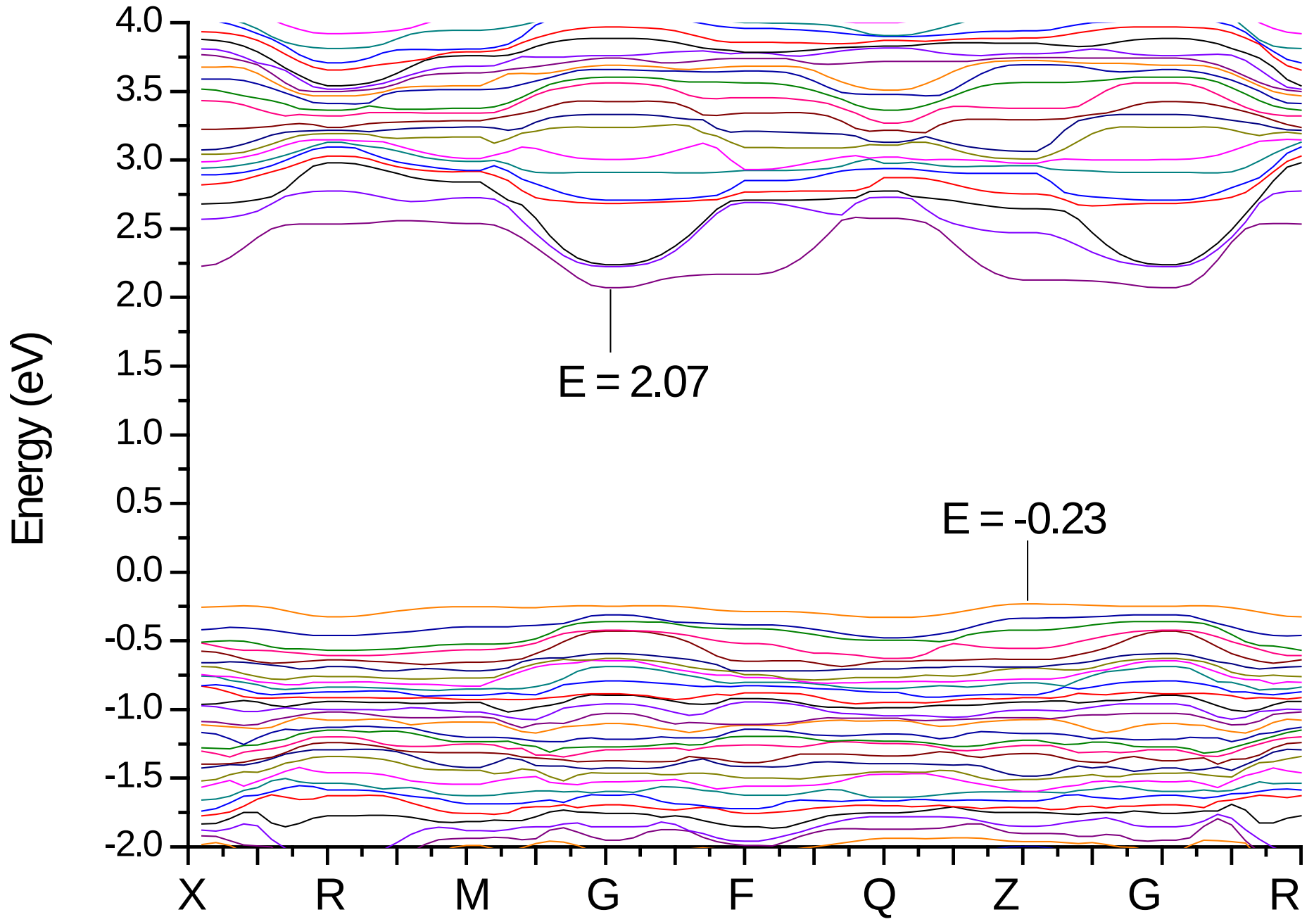

**Figure 4.** Band structure of a 2×2×2 STO cell with one Sr-Ti antisite defect.The indirect energy gap is 2.07 eV – (-0.23 eV) = 2.30 eV.

**Figure 5**

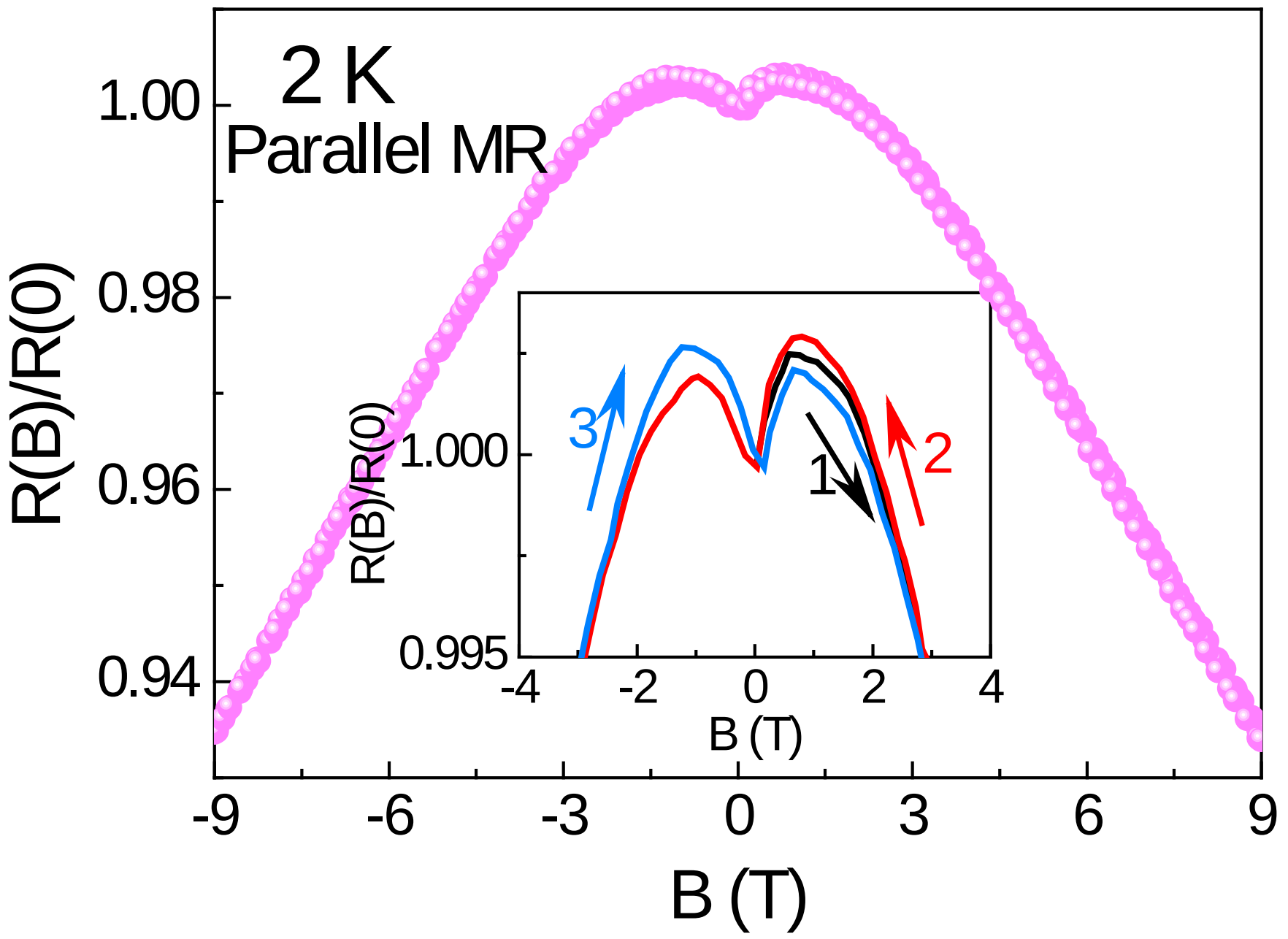

**Figure 5. Parallel MR at 2 K of an aLAO/STO/LAO heterostructure.** Inset: zoom-in MR curves at small fields. The arrows accompanied by numbers represent the measurement sequence during continuous field scans from 0 → 9 T → -9 T → 9 T.

**Figure 6**

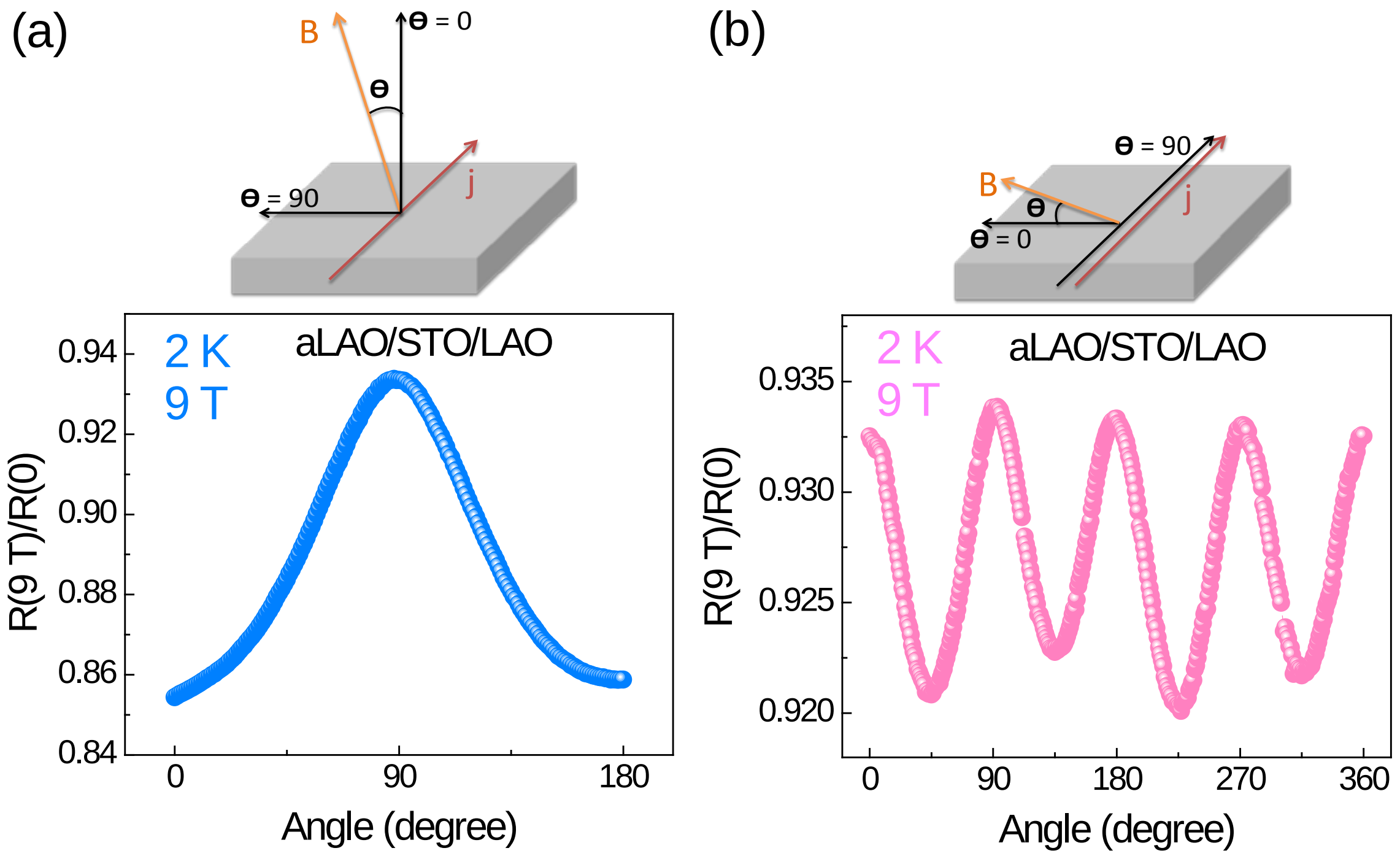

**Figure 6. Anisotropic magnetoresistance (AMR).** (a) Out-of-plane and (b) in-plane AMR of the 2DEG-V in aLAO/STO/LAO at 2 K under a field of 9 T. The measurement geometries are schematized on top of figures. All the resistance is normalized to zero-field resistance.

# Supporting Information for "Bandgap Controlling of the Oxygen-Vacancy-Induced Two-Dimensional Electron Gas in $SrTiO_3$"

Z. Q. Liu[1,2#]*, W. Lu[3], S. W. Zeng[1,2], J. W. Deng[1], Z. Huang[1], C. J. Li[1,4], M. Motapothula[1,5], W. M. Lü[1], L. Sun[1], K. Han[1,2], J. Q. Zhong[2], P. Yang[6], N. N. Bao[1],W. Chen[2], J. S. Chen[3], Y.P. Feng[2], J. M. D. Coey[1,7], T. Venkatesan[1,2,4,8], and Ariando[1,2]*

[1]*NUSNNI-Nanocore, National University of Singapore, 117411 Singapore*

[2]*Department of Physics, National University of Singapore, 117542 Singapore*

[3]*Department of Materials Science and Engineering, National University of Singapore, 117576 Singapore*

[4]*National University of Singapore (NUS) Graduate School for Integrative Sciences and Engineering, 28 Medical Drive, Singapore 117456*

[5]*Center for Ion Beam Applications, National University of Singapore, 117542 Singapore*

[6]*Singapore Synchrotron Light Source (SSLS), National University of Singapore, 5 Research Link, Singapore 117603*

[7]*Department of Pure and Applied Physics, Trinity College, Dublin 2, Ireland*

[8]*Department of Electrical and Computer Engineering, National University of Singapore, 117576 Singapore*

[#]Current address: Oak Ridge National Laboratory, Oak Ridge, Tennessee 37831, United States.

*To whom correspondence should be addressed.
E-mail: liuz@ornl.gov
E-mail: ariando@nus.edu.sg

**S1. XRD patterns of 150-nm-thick STO films deposited on LAO at different temperatures**

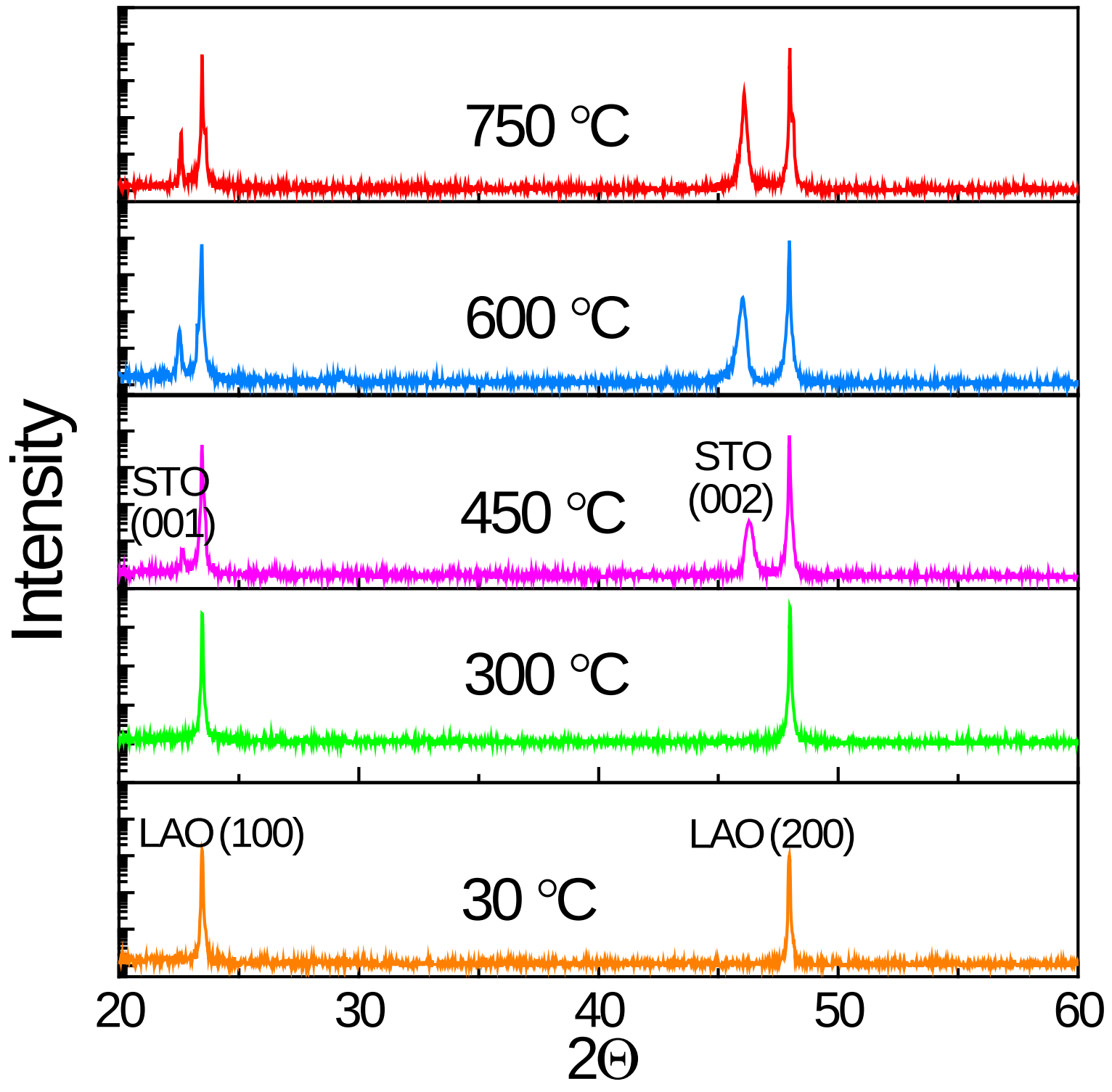

**Figure S1.** X-ray diffraction (XRD) spectra of STO films deposited on LAO substrates at $10^{-2}$ Torr and different temperatures. No detectable XRD peak is present in STO films grown below 450 °C.

## S2. XPS of STO films deposited at $10^{-2}$ Torr and 750 °C

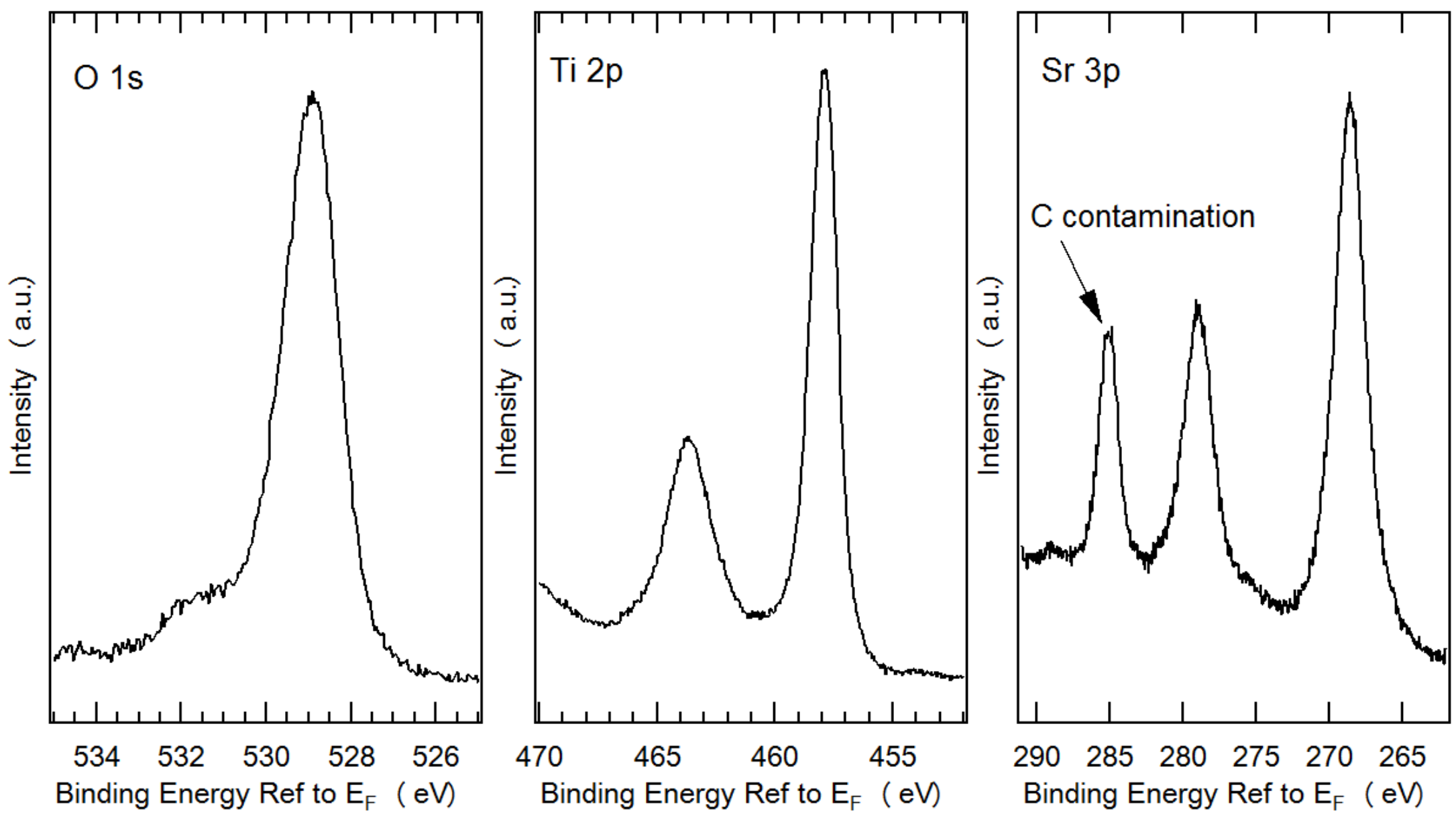

**Figure S2.** XPS spectra of an as-deposited 150-nm-thick STO film fabricated at $10^{-2}$ Torr and 750 °C on LAO. No detectable peak of $Ti^{3+}$ was seen. The charging effect was neutralized by an electron generator and the energy in all the spectra were referred to the C contamination peak-285 eV.

## S3. Consideration on the effect of cationic vacancies on bandgap

(1) As we did not observe any $Ti^{3+}$ from XPS measurements, the true formula in the case of cationic vacancies will be $(SrTi)_{1-x}O_3$. For the sake of charge balance, oxygen charge must be reduced from 2 to 2(1-*x*). This would have several effects:

   a. Lattice parameter is little changed
   b. Ti remains 4+
   c. 2*p*(O) band is no longer full as 2*p* band filling depends on the number of oxygen electrons
   d. $t_{2g}$–$e_g$ crystal field splitting of Ti atoms in the oxygen octahedron is reduced because the crystal field splitting depends on oxygen charge
   e. Consequently, bandgap increases as schematized below.

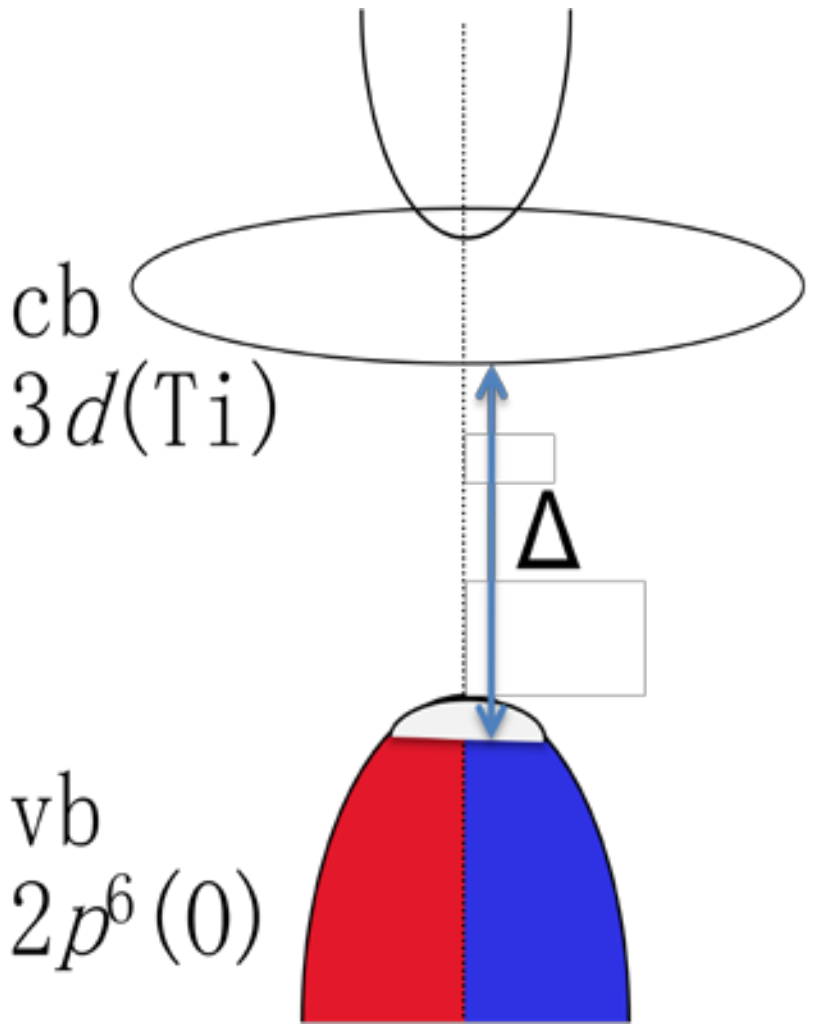

**Figure S3.** Schematic of STO band structure with cationic vacancies.

In this case, an estimated 6% cationic vacancies is expected to generate a bandgap increase of 0.31 eV in crystalline STO films. This corresponds to a reduction in density of 4.4%. We therefore performed x-ray reflectivity for our crystalline STO films. But unfortunately, we found that the density of such crystalline STO films was comparable with that of an idea STO single crystal.

**S4. Synchrotron-based x-ray reflectivity measurement**

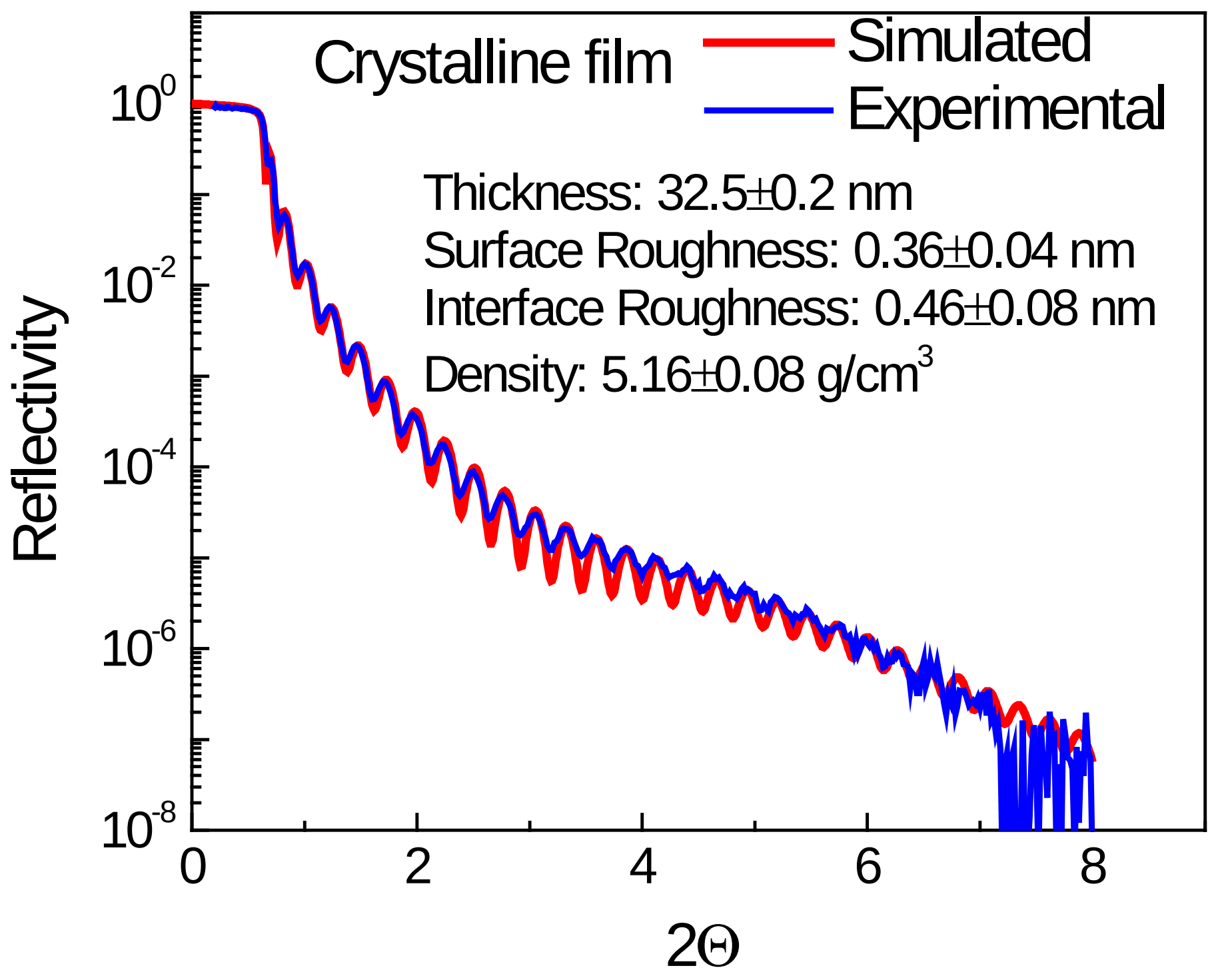

**Figure S4.** X-ray reflectivity (XRR) spectrum of a 32-nm-thick crystalline STO film deposited at $10^{-2}$ Torr and 750 °C on LAO, X-ray wavelength 1.538 Å. The fitted density of the STO film is 5.16±0.08 g/cm$^3$, which is in agreement with the density of an ideal STO single crystal 5.11 g/cm$^3$ within the margin error.

**S5. UV spectra of STO films deposited at *various* oxygen partial pressures**

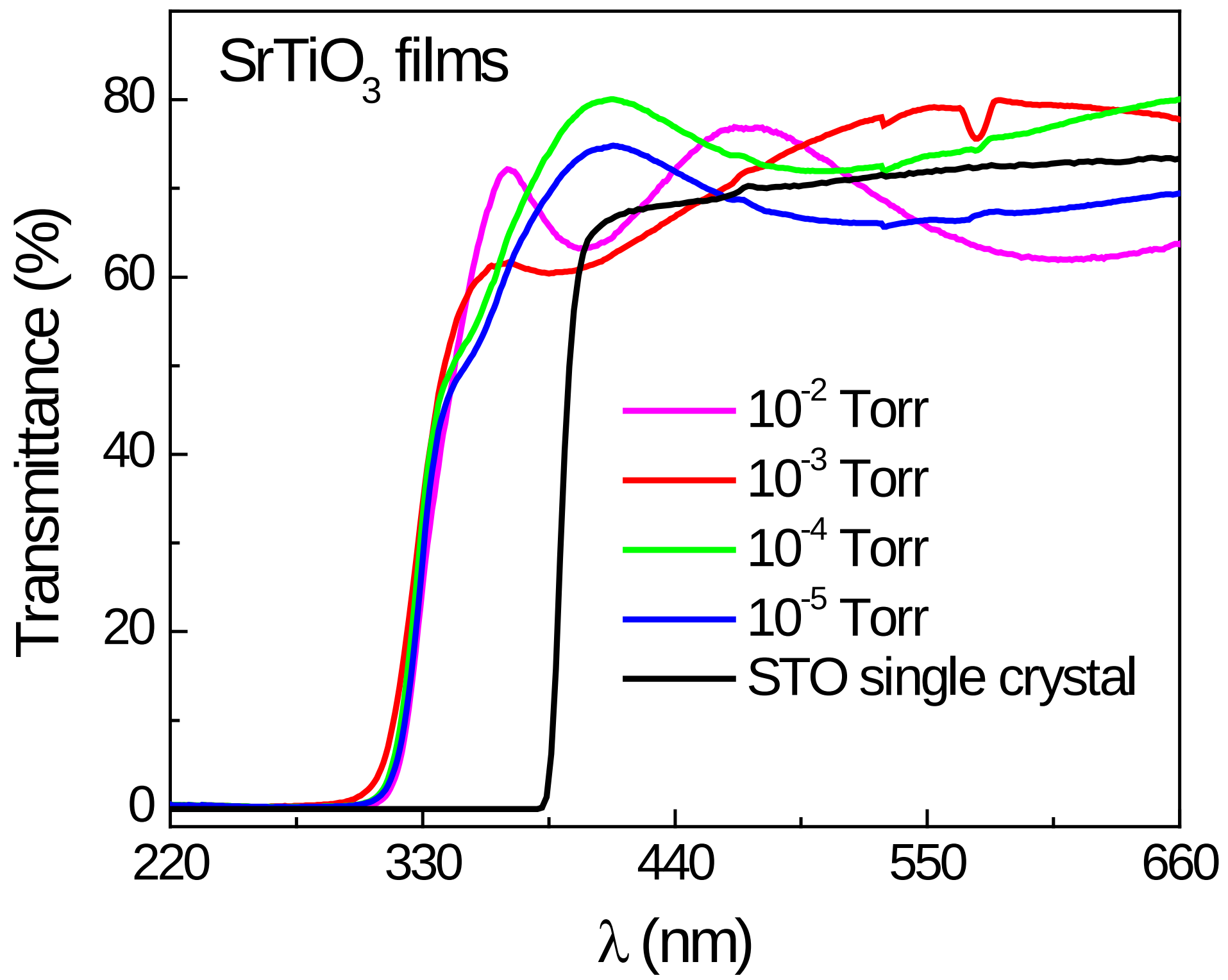

**Figure S5.** UV spectra of STO films deposited on LAO at 750 °C and various oxygen partial pressures.

**S6. RBS of a 150-nm-thick STO film deposited at $10^{-2}$ Torr and 750 °C**

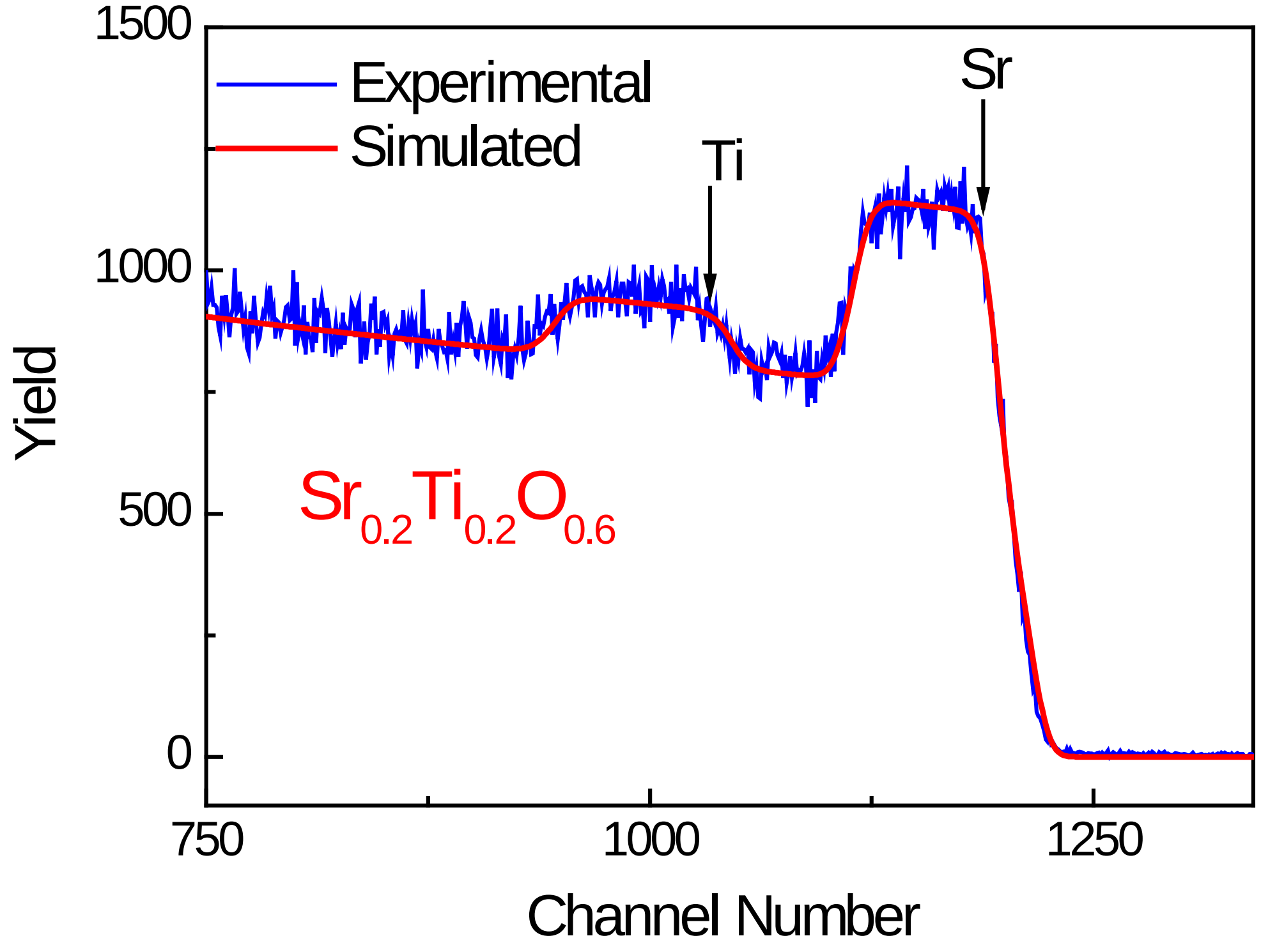

**Figure S6.** RBS spectrum of a 150-nm-thick STO film deposited at $10^{-2}$ Torr and 750 °C on LAO. The simulated composition is $Sr_{0.2}Ti_{0.2}O_{0.6}$.

**S7. UV spectra of STO films deposited by different laser fluence**

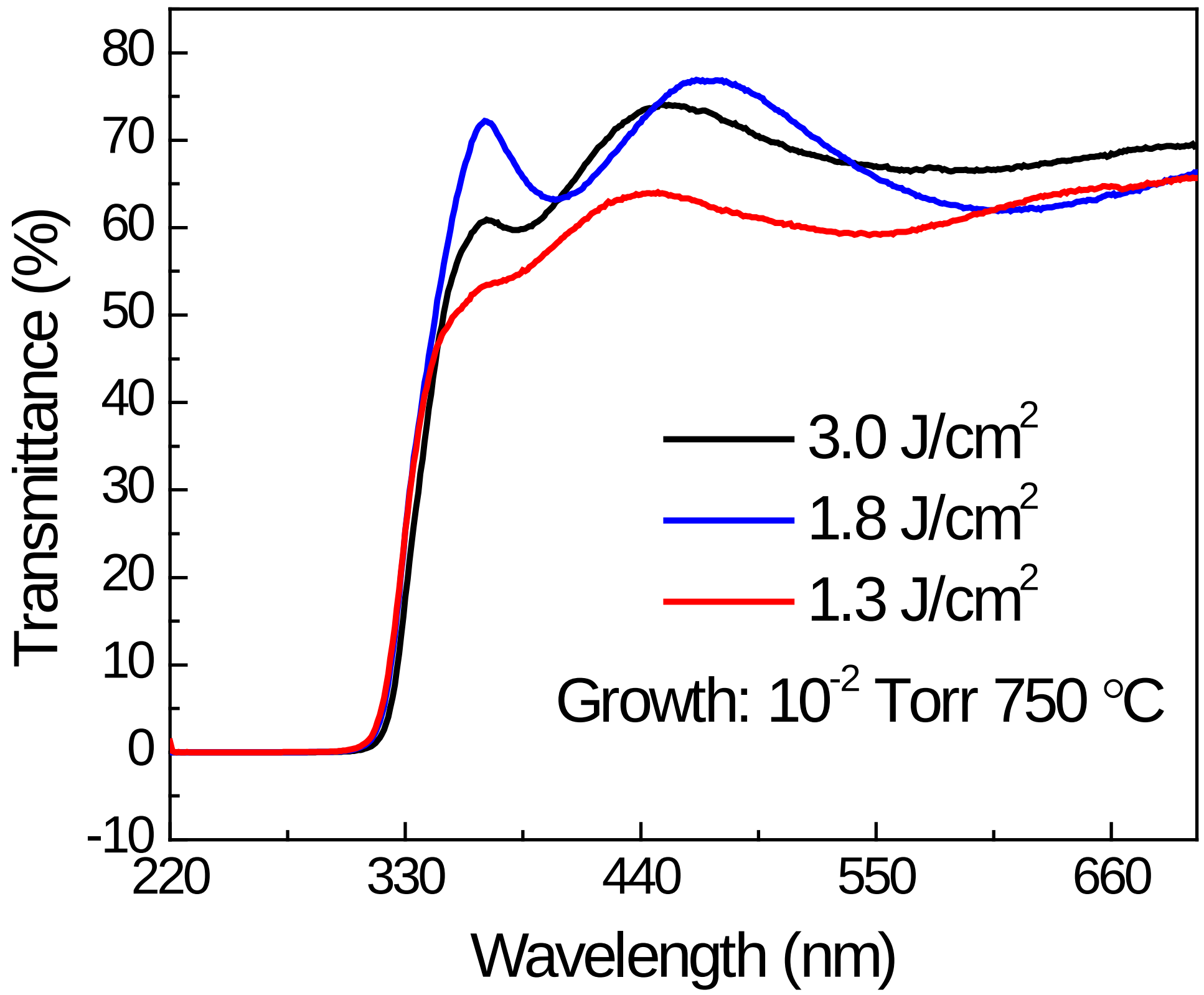

**Figure S7.** UV spectra of STO films deposited on LAO at 750 °C at $10^{-2}$ Torr oxygen partial pressure by different laser fluence.

**S8. Reciprocal space mapping of the STO film**

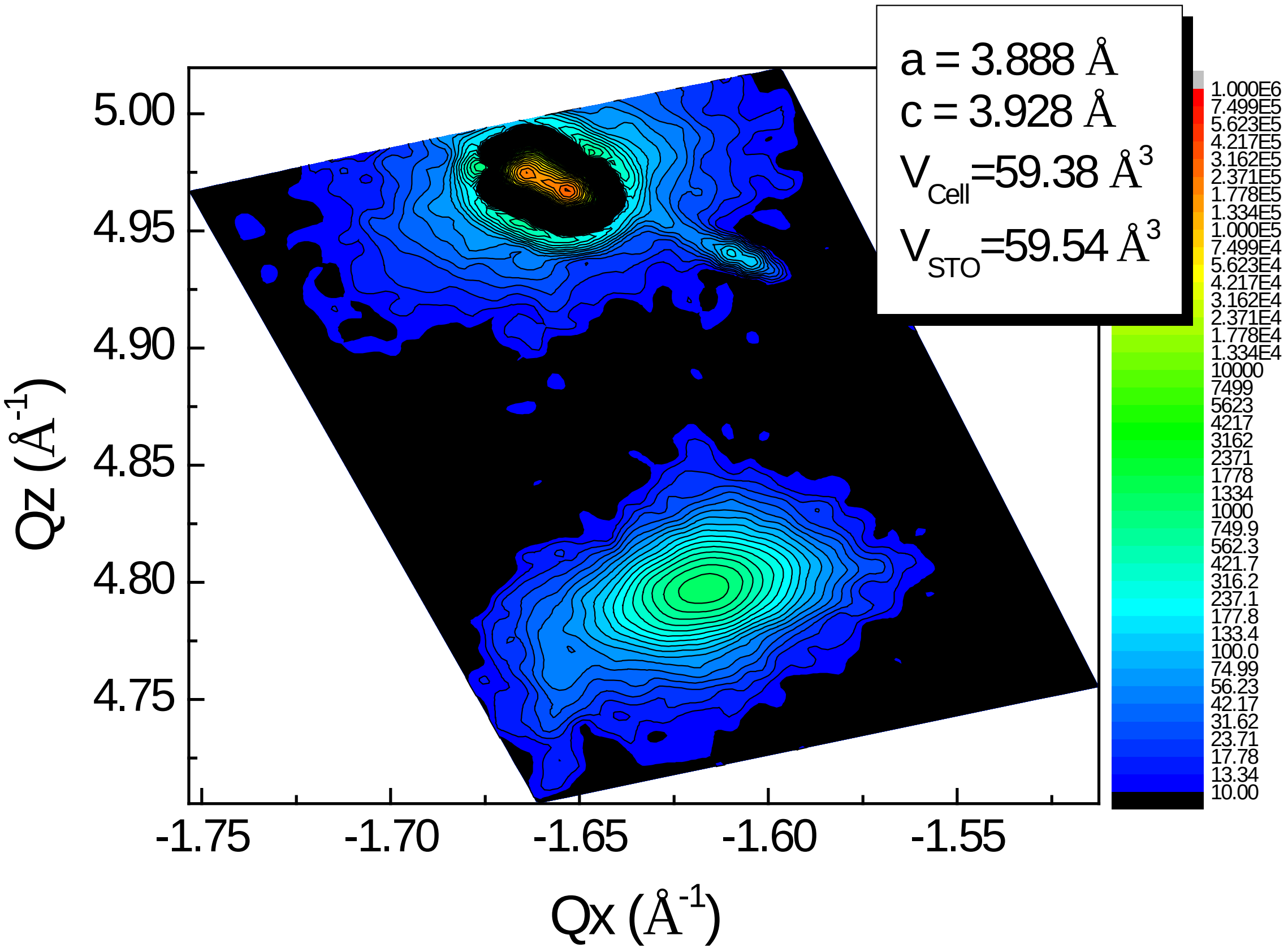

**Figure S8.** (-103) reciprocal space mapping (RSM) of the 32-nm-thick STO film grown on a LAO substrate with the X-ray wavelength 1.538 Å. From the mapping, the film is partially strained. The in-plane lattice constant is calculated to be 3.888 Å and the out-of-plane lattice constant is 3.928 Å. The unit cell volume of the STO film is ~59.38 Å$^3$, which is comparable with that of an ideal STO single crystal 59.54 Å$^3$.

**S9. UV spectra of STO films deposited on *various* substrates**

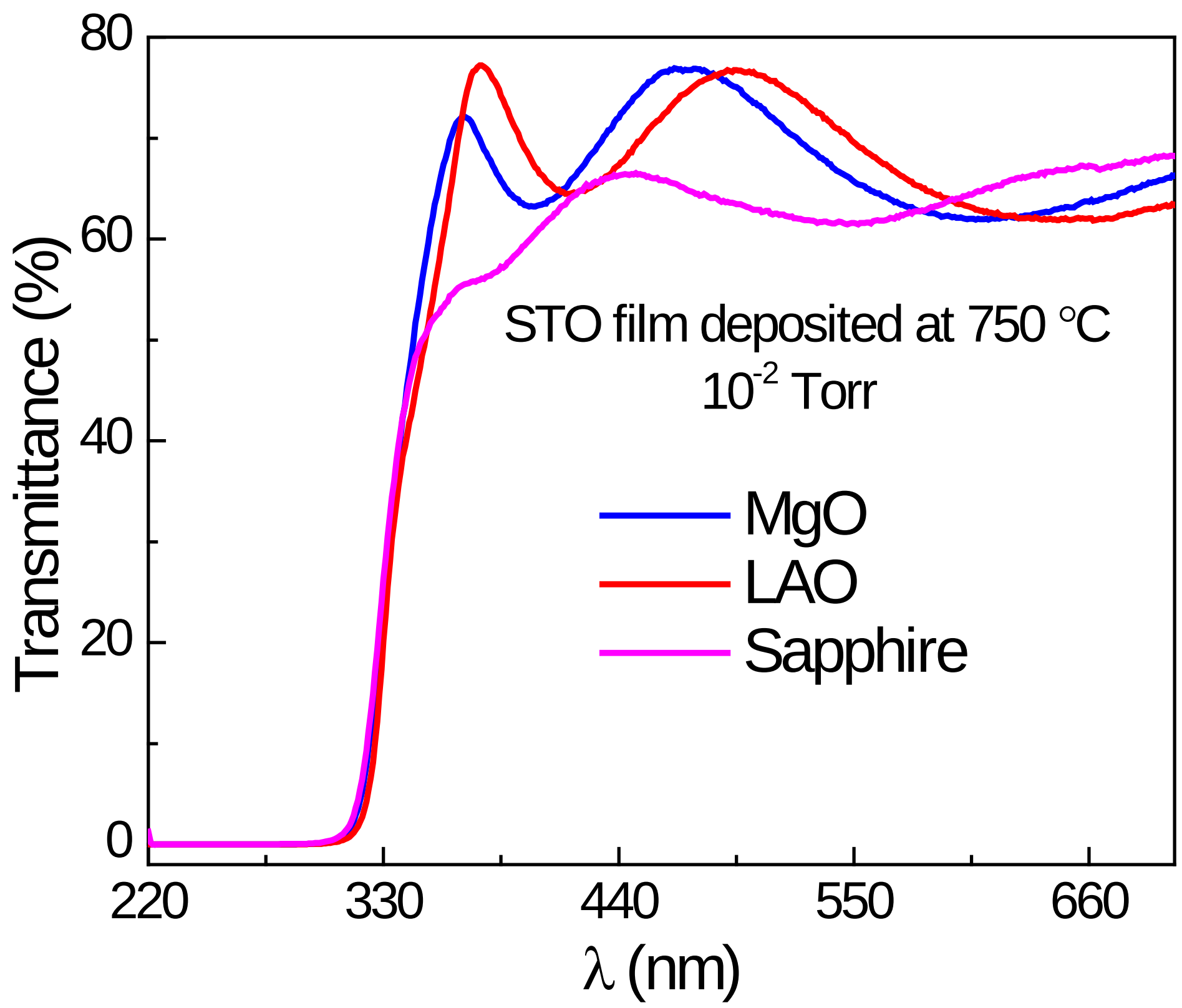

**Figure S9.** UV spectra of STO films deposited on various large bandgap substrates at 750 °C at $10^{-2}$ Torr oxygen partial pressure.

**S10. DFT calculations for the bandgap of single-crystal STO and amorphous STO**

**(1) Single Crystal STO**

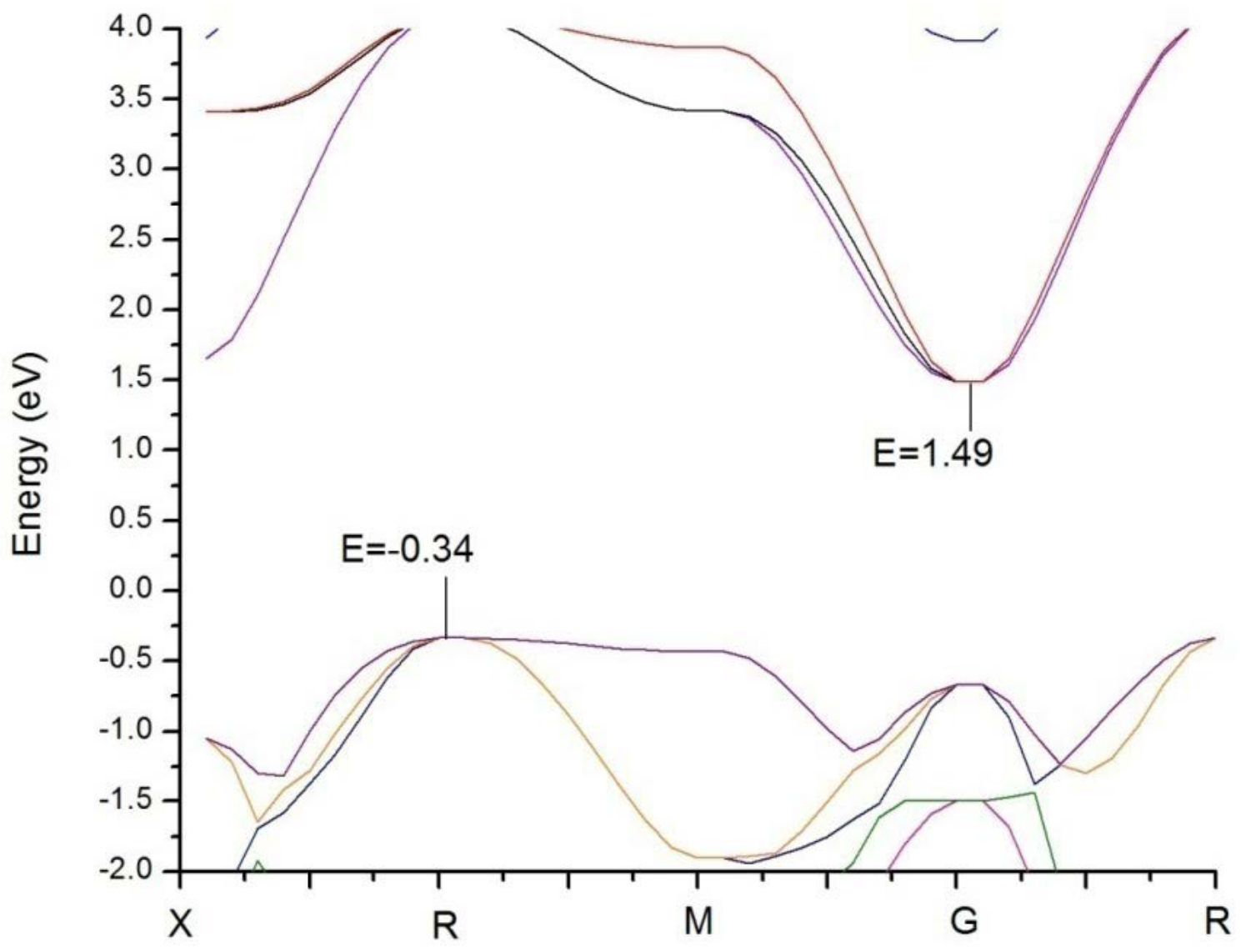

**Figure S10.** Band structure of a STO single crystal.

The calculated bandgap of a STO single crystal is roughly 1.49 eV - (-0.34 eV) ≈ 1.83 eV, comparable with the reported value for the calculated indirect bandgap for STO [van Benthem *et al*. J. Appl. Phys. **90**, 6156 (2001)].

**(2) Amorphous STO**

In our modeling for amorphous STO thin films, we assume the amorphous STO consists of small STO clusters, *i.e.*, we construct different structures of STO as a molecule. Since periodic condition does not apply, there is no real band gap. Instead, we use the excitation energy to explain the transmittance spectrum

For a cluster with 2×2×1 unit cells, with 8 Ti, 9 Sr and 28 O. Following graph shows the energy levels near HOMO and LUMO.

| Energy (eV) | Occupation |
|---|---|
| -4.9294 | 1.99524 |
| -4.8556 | 1.53753 |
| -4.8556 | 1.53753 |
| -4.8267 | 0.93446 |
| -4.6314 | 0.00000 |
| -4.5191 | 0.00000 |
| -1.7692 | 0.00000 |
| -1.6337 | 0.00000 |
| -1.6337 | 0.00000 |

Since our measurement of the bandgap is restricted to UV range, the corresponding excitation energy counts from -4.8 to -1.7 eV, which is roughly 2.9 eV. As mentioned above, the cluster no longer is treated as an isolated molecule, and the excitation energy in UV range should be responsible for the low transmittance in our experiment (Figure XX). In other words, the excitation energy in our modeling acts similarly with the usual band gap in periodic sttructure.

For a cluster with dimensions 3×2×1 unit cells, with 12 Ti, 12 Sr and 40 O:

| Energy (eV) | Occupation |
|---|---|
| -5.2112 | 1.98378 |
| -5.1979 | 1.95752 |
| -5.1830 | 1.89201 |
| -5.1827 | 1.89037 |
| -5.0884 | 0.28586 |
| -5.0249 | 0.00416 |
| -4.8261 | 0.00000 |
| -4.7929 | 0.00000 |
| -2.1722 | 0.00000 |

The excitation energy is -2.2 eV - (-5.1 eV) ≈ 2.9 eV.

For a more spherical cluster, with 8 Ti, 19 Sr and 36 O. The cluster is schematized as below:

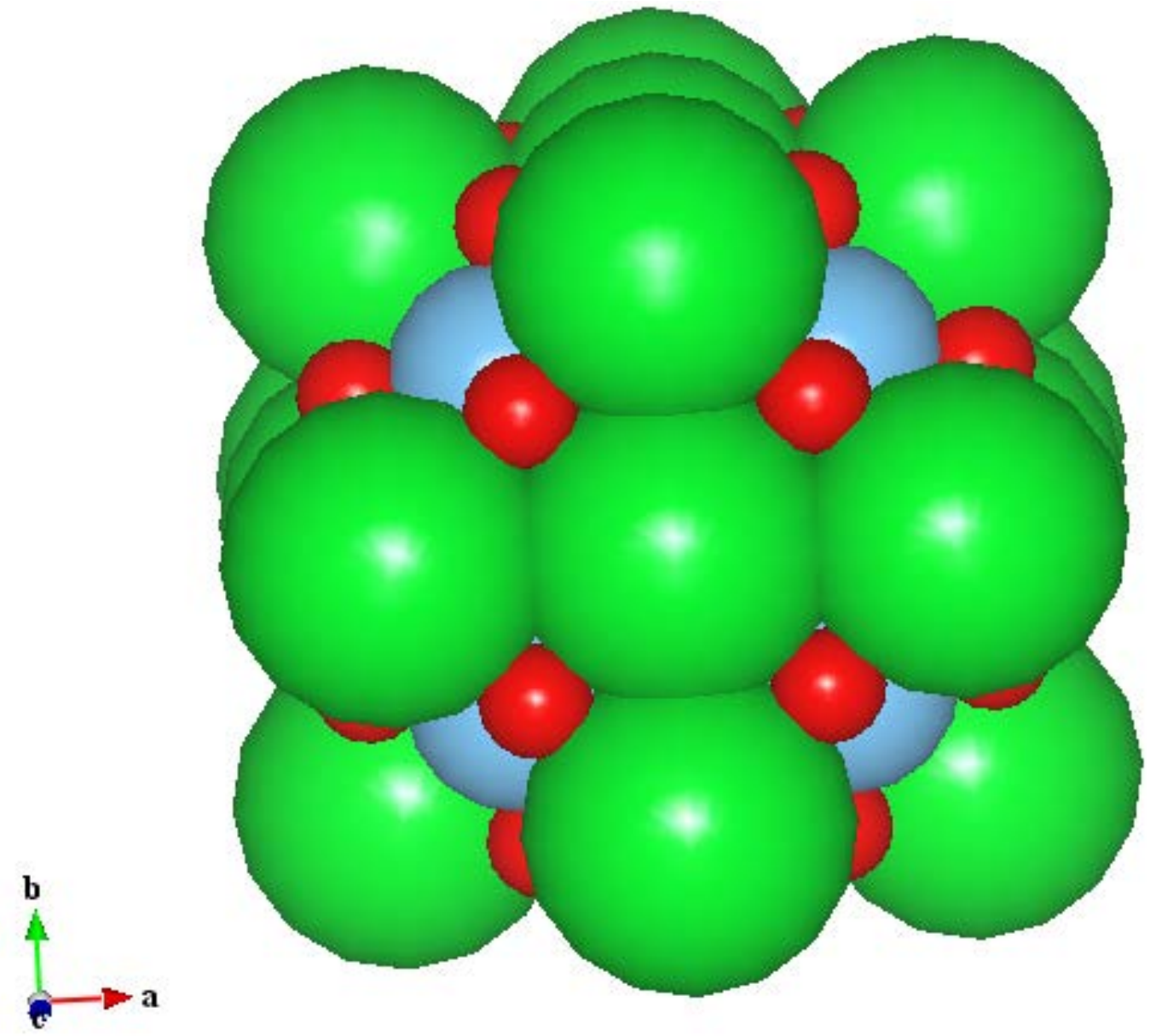

**Figure S11.** Schematic of the cluster model of amorphous STO.

The occupations are

| Energy (eV) | Occupation |
|---|---|
| -3.3999 | 1.99927 |
| -3.3999 | 1.99926 |
| -3.3276 | 1.81612 |
| -3.2847 | 1.09399 |
| -3.2847 | 0.28586 |
| -0.5393 | 0.00000 |
| -0.3541 | 0.00000 |
| -0.3541 | 0.00000 |
| -0.3540 | 0.00000 |

The - excitation is -0.5 eV - (-3.3 eV) ≈ 2.8 eV.

**S11. DFT calculations of the STO bandgap with cationic vacancies**

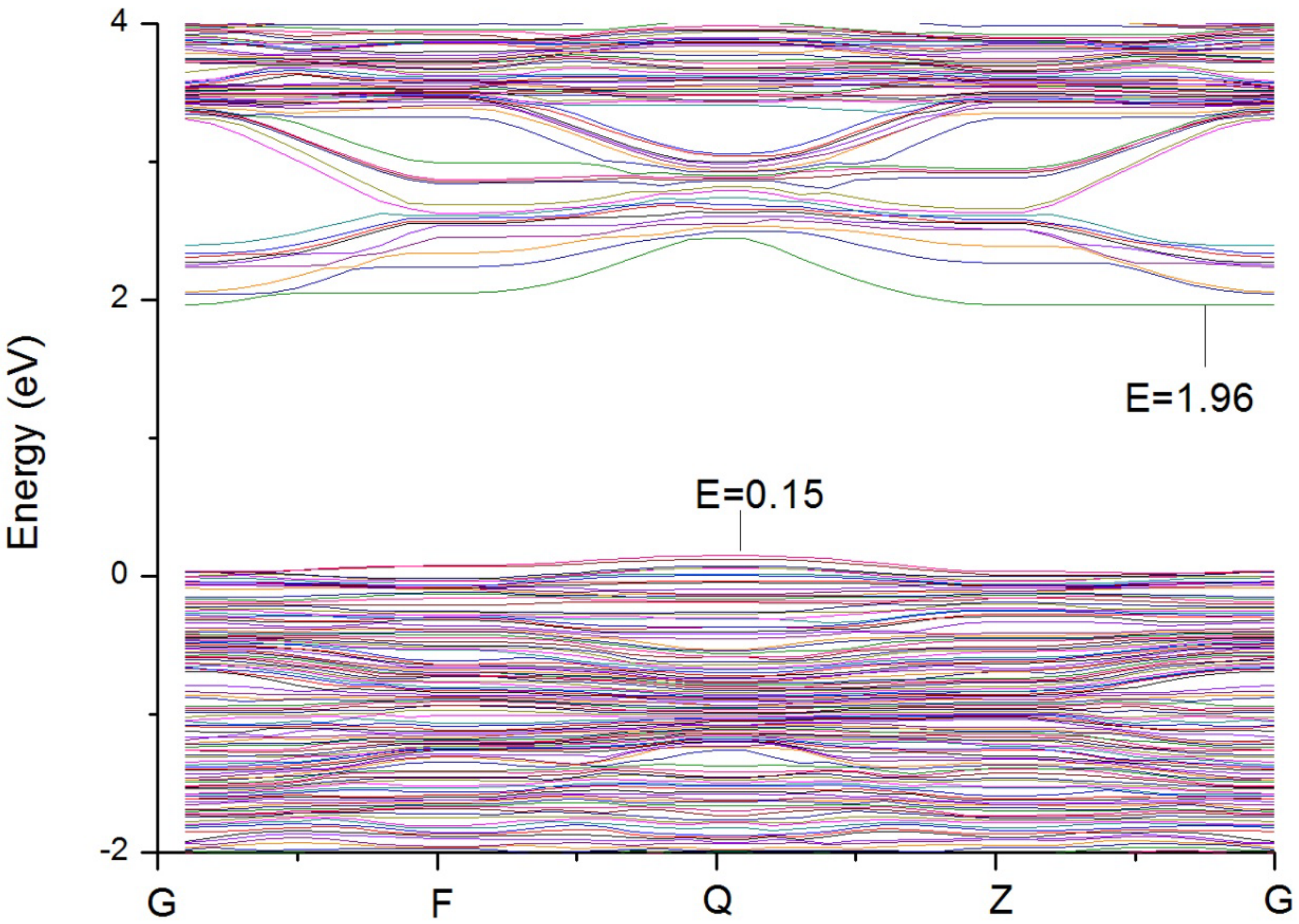

**Figure S12** Band structure of a 3×3×3 STO cell with 2 Ti and 2 Sr vacancies. The energy gap is 1.96 eV – 0.15 eV = 1.81 eV, which is comparable with the calculated bandgap of an ideal STO single crystal. Regarding Sr or Ti vacancies, the bandgap enlargement is thus not pronounced.

**S12. Magnetoresistance data of aLAO/STO/LAO and aLAO/STO heterostructures**

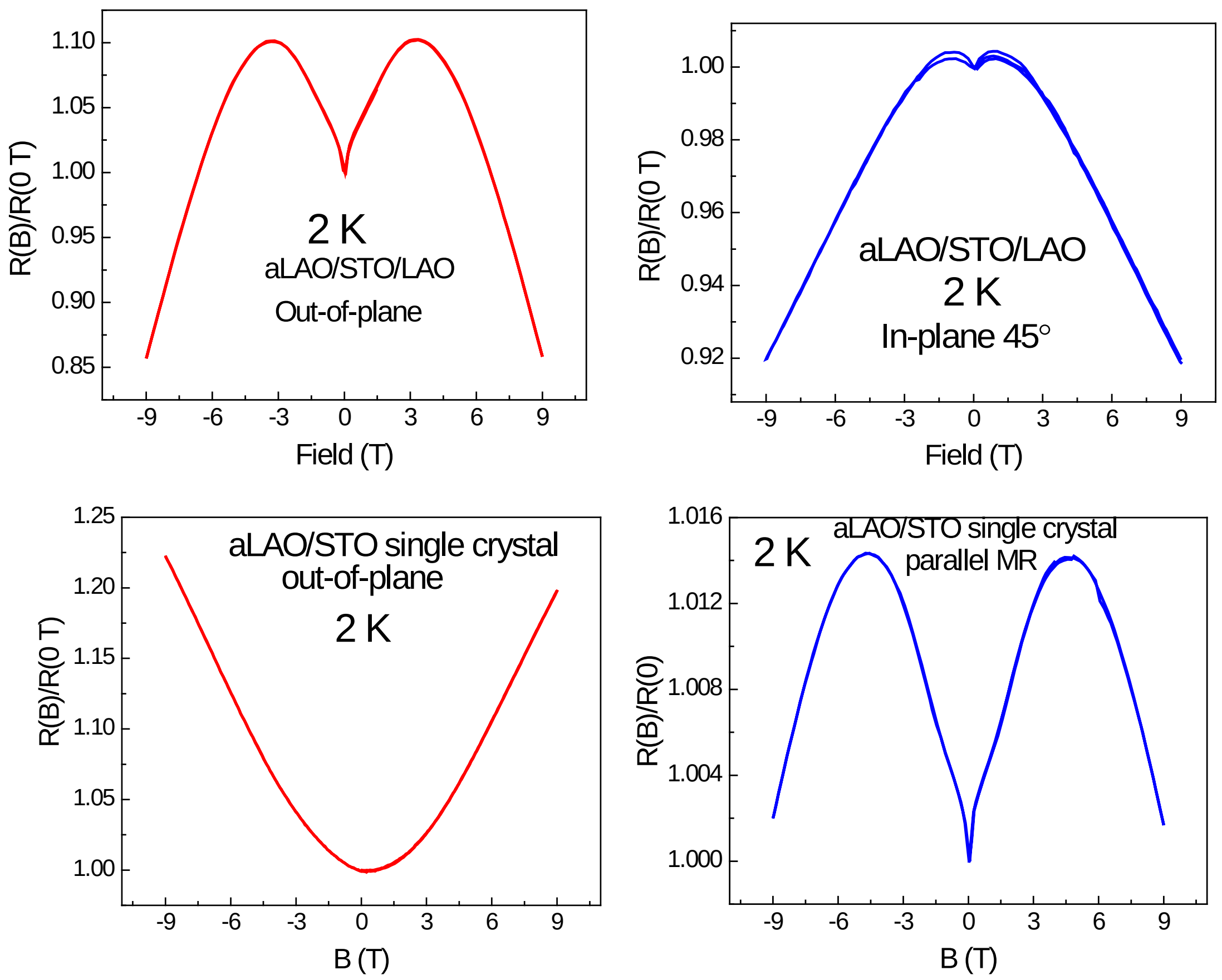

**Figure S13** MR data of aLAO/STO/LAO (STO layer deposited at 750 °C) and aLAO/STO heterostructures with different measurement geometries.